\documentclass[]{raa}            
\usepackage{graphicx,times}
\usepackage{natbib}

\providecommand{\bm}[1]{\mbox{\boldmath$#1$\unboldmath}}

\begin{document}

  \title{Nonlinear time series anaysis of the light curves from the black hole system GRS1915+105
}

 \volnopage{ {\bf 2010} Vol.\ {\bf X} No. {\bf XX}, 000--000}
   \setcounter{page}{1}

   \author{K. P. Harikrishnan
      \inst{1}
   \and R. Misra
      \inst{2}
   \and G. Ambika
      \inst{3}
   }

   \institute{Department of Physics, The Cochin College, Cochin-682002,
             India; {\it $kp_{\_}hk2002@yahoo.co.in$}\\
        \and
             Inter University Centre for Astronomy and Astrophysics,
              Pune-411007, India; {\it rmisra@iucaa.ernet.in}\\
        \and
             Indian Institute of Science Education and Research,
              Pune-411021, India; {\it g.ambika@iiserpune.ac.in}\\
\vs \no
   {\small Received [year] [month] [day]; accepted [year] [month] [day] }
}

\abstract{GRS 1915+105 is a prominent black hole system exhibiting variability over a wide range of 
time scales and its observed light curves have been classified into 12
temporal states. Here we undertake a complete analysis of these light curves from all the 
states using various quantifiers from nonlinear time series analysis, such as, the correlation 
dimension ($D_2$), the correlation entropy ($K_2$), singular value decomposition (SVD)and 
the multifractal spectrum ($f(\alpha)$ spectrum). An important aspect of our analysis is 
that, for estimating these quantifiers, we use algorithmic schemes which we have 
proposed recently and tested successfully on synthetic as well as practical time series 
from various fields. 
Though the schemes are based on the 
conventional delay embedding technique, they are automated so that the above 
quantitative measures can be computed using conditions prescribed by the algorithm 
and without any intermediate subjective analysis. We show that 
nearly half of the 12 temporal states exhibit deviation from randomness and their complex 
temporal behavior could be approximated by a few (3 or 4) coupled ordinary nonlinear 
differential equations. These results could be important for a better understanding 
of the processes that generate the light curves and hence for modelling the temporal 
behavior of such complex systems. To our knowledge, this is the first complete analysis of an 
astrophysical object (let alone a black hole system) using various techniques from 
nonlinear dynamics. 
\keywords{accretion, accretion disks: X-rays: binaries
}
}

   \authorrunning{K. P. Harikrishnan, R. Misra \& G. Ambika }            
   \titlerunning{Nonlinear analysis of black hole light curves }  
   \maketitle

\section{Introduction}
\label{sect:intro}
Most of the systems in Nature are described by models which are inherently nonlinear. 
Since the discovery of  {\it deterministic chaos} a few decades back and the development 
of various techniques in subsequent years, there remained the exciting prospect of a 
better understanding of the complex behavior shown by various natural systems in terms of 
simple nonlinear models. Evidence for low dimensional chaos has been reported - and 
disputed - not only in physical sciences, but also in many other fields such as, 
physiology, economics and social sciences \citep{sch1}. 
Particular attention has been paid 
to systems producing strange and chaotic attractors, the word {\it strange} refering to 
metric properties such as fractal dimension and the word {\it chaotic} representing dynamic 
properties like exponential divergence of nearby trajectories in phase space. A large number 
of techniques and measures from nonlinear dynamics and chaos theory are routinely being 
employed for the analysis of such systems. Excellent text books are now available that 
give a background knowledge on various methods in nonlinear dynamics 
\citep{hil,spr,lak}. 

One major difficulty in the analysis of real world systems is that our knowledge regarding 
the system is usually limited to a single scalar variable recorded as a function of time, 
called the {\it time series}. Therefore, a great deal of effort has been devoted to the 
characterisation of underlying attractors reconstructed from time series. The large number of 
techniques and computational schemes used for this purpose have been discussed in detail 
by many authors \citep{kan1,abe,heg1}.

Among the most important quantifiers used for the analysis of time series data are the 
correlation dimension ($D_2$), the correlation entropy ($K_2$) and the multifractal spectrum. 
Correlation dimension is often used as a discriminating statistic for hypothesis testing 
to detect nontrivial structures in the time series. But when the time series involves 
colored noise, a better discriminating measure is considered to be $K_2$  
\citep{ken}. 
Finally, a complete characterisation of the underlying chaotic attractor is done using 
the generalised dimensions $D_q$ and the $f(\alpha)$ spectrum.

We have recently proposed automated algorithmic schemes \citep{kph1,kph2}
for the computation of 
$D_2$ and $K_2$ from time series based on the delay embedding technique and applied it 
successfully to various types of time series data including that from standard chaotic systems, 
data added with white and colored noise and practcal time series like EEG and ECG. 
A generalisation of these schemes to compute the multifractal spectrum of a chaotic attractor 
has also been proposed \citep{kph3,kph4}. These schemes provide a 
nonsubjective approach for the 
characterisation of strange attractors inherent in time series.

It should be noted that so far, most of the analysis of the light curves from X-ray binaries 
and active galactic nuclei (AGN) have used the conventional techniques such as the power 
spectrum and distribution. It is widely believed that the light intensity variations are mostly 
stochastic in nature. For example, it has been shown in the case of the most prominent black hole 
system, 
Cygnus X-1, that the observed light curves, at least in certain time scales, are consistent with 
some static nonlinear transformatons of stochastic variations in intensity 
\citep{utt}. 
The authors argue that models based on nonlinear dynamics are not required to explain the data. 
 
But there are also some  analysis based on nonlinearity measures that have been attempted 
earlier \citep{vog,nor,tim} on the 
light curves of some prominent black hole systems, such as, 
Her X-1 and Cygnus X-1. But these studies have so far not been able to provide conclusive 
evidence for nontrivial structures in the temporal behavior of such systems. One reason for this 
has been the limited number of data sets available from such sources with sufficient signal to 
noise ratio required for such analysis \citep{nor}. The scenario has changed in the last few years 
as enough data are now available through RXTE observations. 
Recently, a nonlinear time series analysis performed on light intensity data from the 
white dwarf variable PG1351+489 has enabled much more information regarding the system 
compared to the conventional power spectrum analysis \citep{jev} and attempts have been
made to use these analysis to differentiate between neutron stars and black holes \citep{Kar10}

Studies on 
GRS1915+105 have been limited because it became active only over a decade ago. But the system turns 
out to be  unique among all such sources in that it seems to flip from one state to another 
continuously with each state having its own temporal variability different from the other 
states. The light curves have been classified into 12 spectroscopic classes based on 
RXTE observations by \citet{bel}. The nature of the light curves changes completely as the system flips  from 
one state to another. Evidently, pure stochastic processes cannot account for such qualitative 
changes in the light curves. Hence the question naturally arises whether some nonlinear 
deterministic processes are also involved. We do find evidence to support this.

In this paper, we apply the above mentioned automated  schemes developed by us to undertake a 
complete analysis of the X-ray light 
curves from GRS1915+105. Earlier, a surrogate analysis with $D_2$ as the 
discriminating measure has 
shown that a few of these states manifest the time evolutions analogous to that from low 
dimensional nonlinear systems with some inherent noise \citep{mis1}. 
This motivates us to 
undertake an exhaustive numerical analysis of the light curves from the source in all its 
temporal states using the prominent tools from nonlinear dynamics.

Another motivation for the 
present investigation has been derived from the fact that the accretion disk in such systems 
are driven by
magneto-hydrodynamic turbulence which is an intrinsically non-linear process.
A model for such a process should be nonlinear and is expected to  
show qualitative changes in its behavior as a control parameter is varied. For the X-ray 
radiation from an accretion disk, the rate of mass accretion could possibly be considered 
as a  suitable control parameter. There also exists theoretical models to this effect 
\citep{vog,atm1} from which it is possible to derive 
the temporal variability of the X-ray 
radiation in different regimes of mass accretion rate. Since many of the states of the 
black hole system under study show nonlinear character, a complete 
analysis of the light curves using various nonlinear measures can greatly help in the search for 
a nonlinear deterministic model to describe the temporal variability of the system.

Our paper is organised as follows: All the quantitative measures used in this paper and the 
corresponding computational schemes are discussed in detail in the following  section. 
While \S 2.1 and \S 2.2 present the computational details for $D_2$ and $K_2$, \S 2.3  
and 2.4 are concerned with SVD and $f(\alpha)$ spectrum respectively. The time series from a 
standard chaotic system - the Rossler system - is used as an example to illustrate the 
results in all the cases. The analysis of the X-ray light curves from the black hole system 
is then undertaken in \S 3 and the conclusions are drawn in \S 4.

\begin{figure}
\centering
\includegraphics[width=95mm]{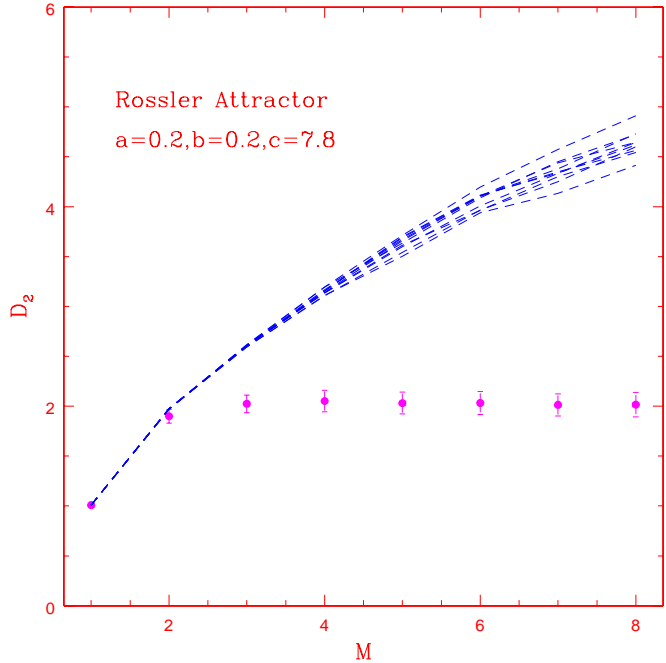}
\caption{The $D_2$ values of the Rossler attractor (with error bar), as a 
function of $M$ along with the $D_2$ values of $20$ surrogates (dashed lines). All 
computations are done with $10000$ data points.}
 \label{Fig1}
\end{figure}
 
\section{Quantitative Measures used for the Analysis}
\label{sect:quant}

\subsection{Correlation dimension and surrogate analysis}
Correlation dimension $D_2$ is often used as a discriminating statistic for hypothesis testing. 
The conventional method for the calculation of $D_2$ is the delay embedding method first 
introduced by \citet{tak}, and used effectively by \citet{gra1}, 
now known as the GP algorithm. More details can be found in \citet{sau}. 
It creats an embedding space of dimension $M$ with delay vectors constructed by splitting 
a discretely sampled scalar time series $s(t_i)$ with delay time $\tau$ as 
\begin{equation}
   \label{e.1}
   \vec{x_i} = [s(t_i),s(t_i+\tau),....,s(t_i+(M-1)\tau)]
\end{equation}
The correlation sum is the relative number of points
within a distance R from a particular ($i^{th}$) data point,
\begin{equation}
   \label{e.2}
   p_{i}(R) = \lim_{N_v \rightarrow \infty} {1\over N_v} \sum_{j=1,j 
\neq i}^{N_v} H(R-|\vec{x_{i}}-\vec{x_{j}}|)
\end{equation}
where $N_v$ is the total number of reconstructed vectors and $H$ is the 
Heaviside step function. Averaging this quantity over $N_{c}$ randomly 
selected $\vec{x_i}$ or centers gives the correlation function
\begin{equation}
   \label{e.3}
   C_{M}(R) =  {1\over N_{c}} \sum_{i}^{N_{c}} p_{i}(R)
\end{equation}
The correlation dimension $D_{2}(M)$ is then defined to be,
\begin{equation}
   \label{e.4}
    D_{2} \equiv \lim_{R \rightarrow 0} d ({log} C_M (R))/d ({log} (R))  
\end{equation}
which is the scaling index of the variation of $C_{M}(R)$ with $R$ as 
$R \rightarrow 0$.
In pratice, a linear part in the $log C_M (R)$ versus $log R$ plot is identified 
subjectively, called the scaling region, and its slope is taken as $D_2$. But in our 
computational scheme, this is done algorithmically as discussed in detail 
elsewhere (~\cite{kph1}) and the scheme computes $D_2$ with error bar as a function of $M$. 
The scheme has also been shown to be suitable for hypothesis testing using surrogate data.
 
\begin{figure}
\centering
\includegraphics[width=95mm]{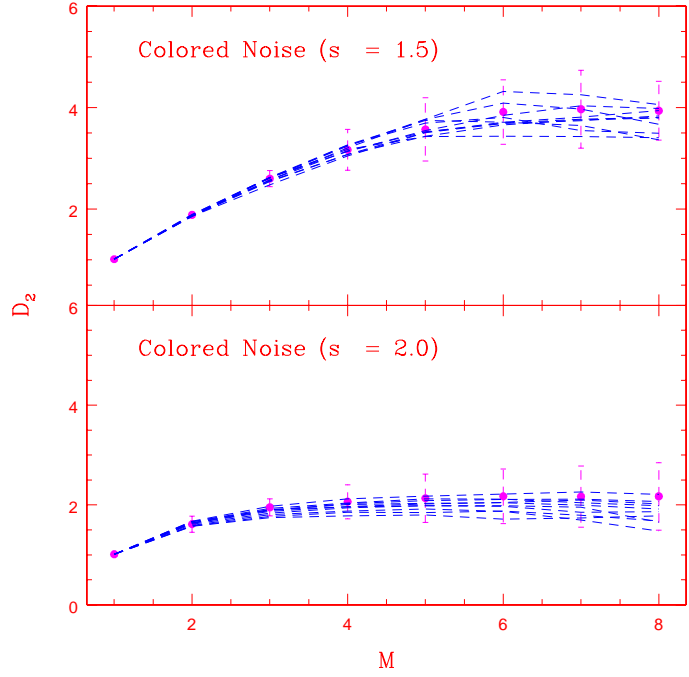}
\caption{The upper panel shows the $D_2$ values of pure colored noise with 
spectral index $s=1.5$ as a function of $M$ along with surrogates. The lower panel shows 
the same for $s=2.0$.}
 \label{Fig2}
\end{figure}

\begin{figure}
\centering
\includegraphics[width=95mm]{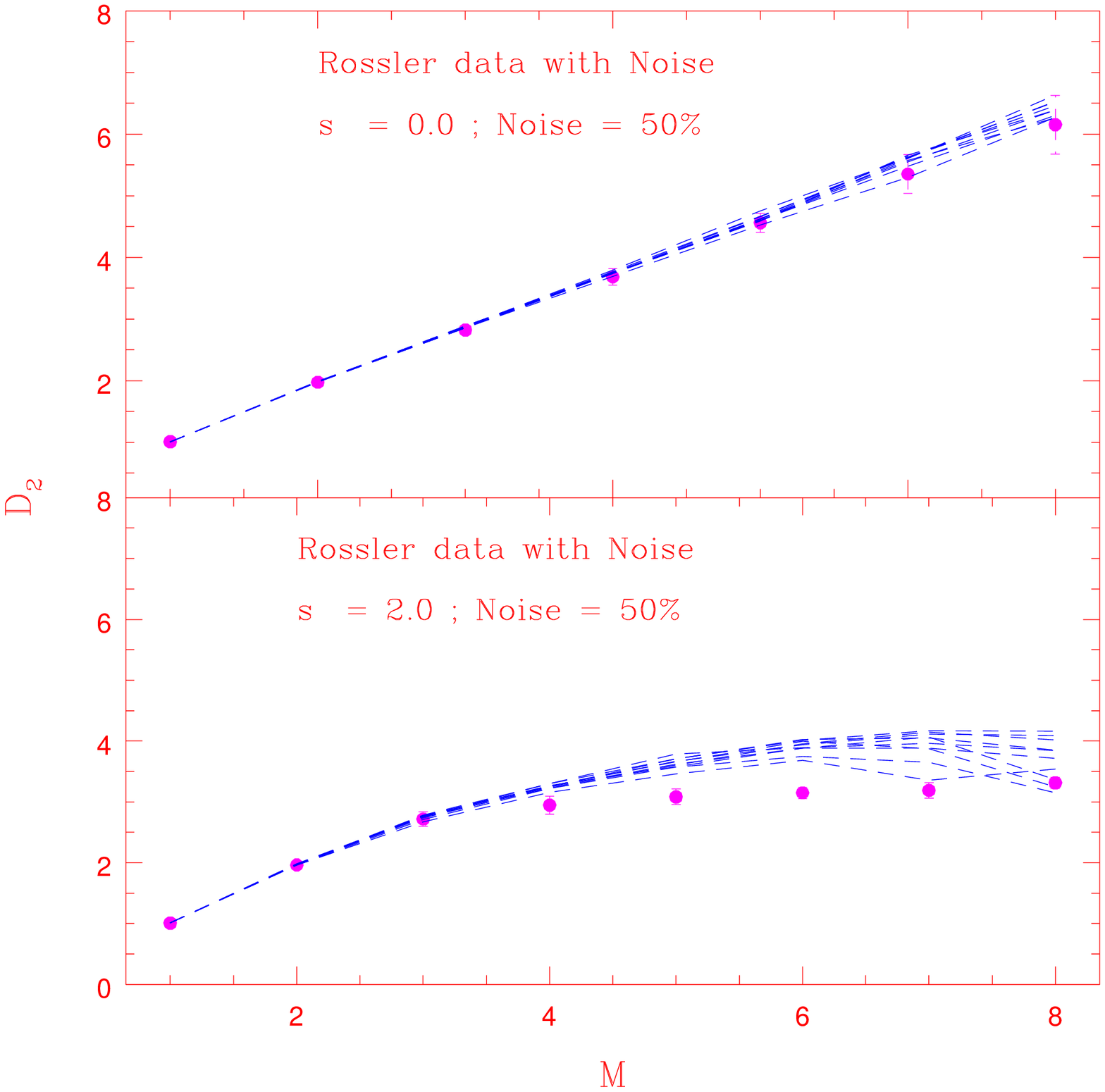}
\caption{The figure shows the $D_2$ values as a function of $M$ for the Rossler 
attractor data added with $50\%$ white noise ($s=0.0$) and the same percentage of colored 
noise ($s=2.0$) along with their respective surrogates. Note that, in the case of colored 
noise contamination (lower panel), the data values are still below that of the surrogates.}
 \label{Fig3}
\end{figure}

The rationale behind surrogate analysis is to formulate a null hypothesis that the data 
has been generated by a stationary linear stochastic process, and then attempt to reject it 
by comparing a suitable measure for the data with appropriate realisations of surrogate data. 
The method for the generation of surrogate data was originally proposed by Theiler and 
coworkers \citep{the1} with the Amplitude Adjusted Fourier Transform (AAFT) algorithm. 
But \citet{sch2,sch3} have proposed another iterative scheme, known as 
IAAFT scheme, which is similar but reported to be more consistent in representing null 
hypothesis \citep{kug} for a wide class of stochastic processes. In this work, we apply 
this scheme to generate surrogate data sets using the TISEAN package \citep{heg1}.

We first apply the $D_2$ analysis on different types of data sets such as, standard chaotic 
time series, pure noise and chaotic data added with noise. The time series from the standard 
Rossler attractor, with parameter values $a=b=0.2$ and $c=7.8$,  
is used as a reference to test all the computational schemes presented in 
this work. All computations are done with $10000$ long data points and $20$ surrogates for 
each data. In Fig.1, $D_2$ of Rossler attractor and surrogates are computed as a 
function of the embedding dimension $M$, where as in Fig.2, the same is shown for two 
pure colored noise data sets with spectral index $s=1.5$ and $2.0$. As expected, the Rossler data 
show clear deviation from the surrogates while for the latter, the null hypothesis cannot be 
rejected. 

\begin{figure}
\centering
\includegraphics[width=95mm]{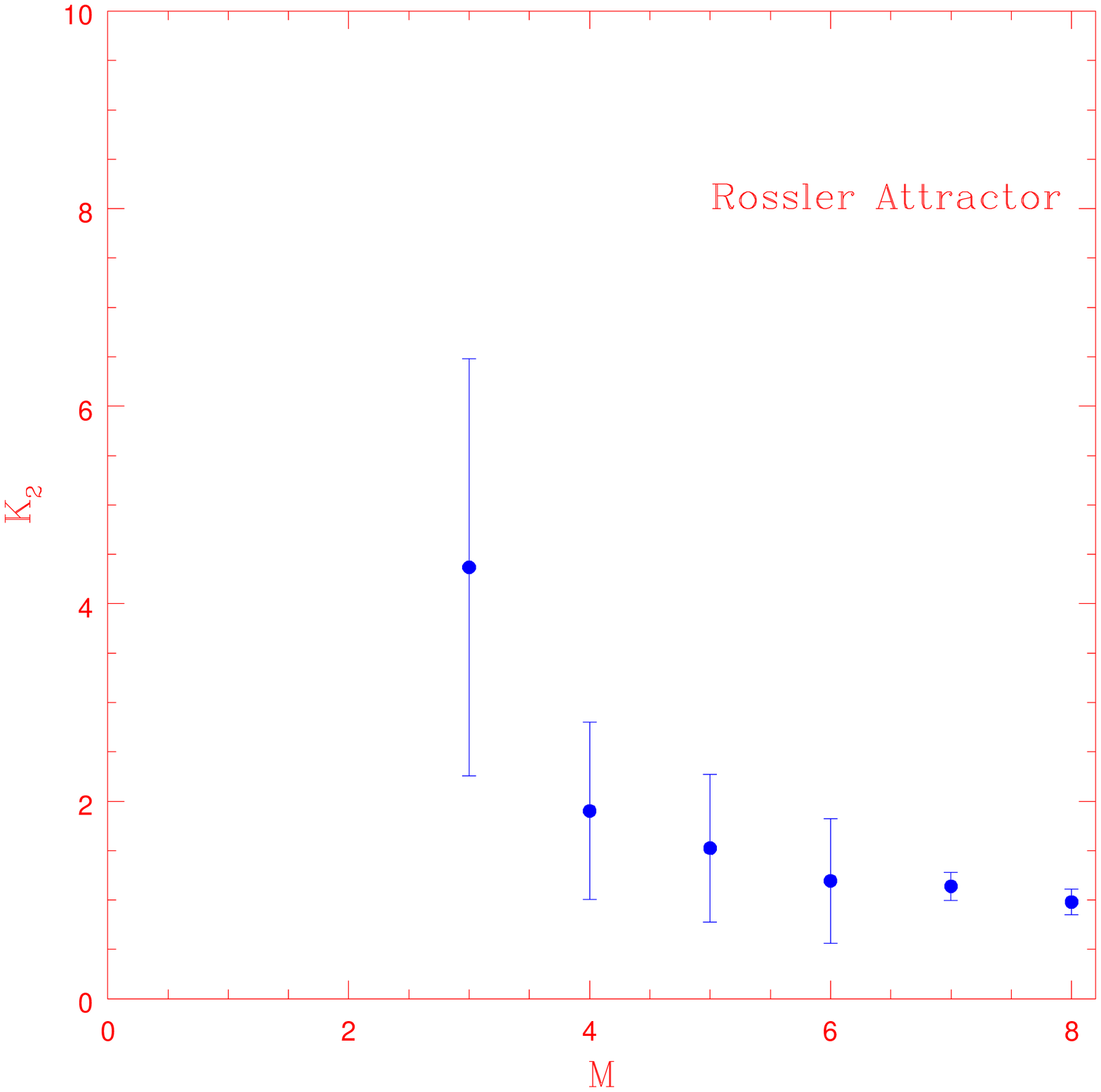}
\caption{The $K_2$ values of the Rossler attractor as a function of $M$ computed 
using our scheme. The values are computed per sec and converges very close to the 
standard value $1.04 \pm 0.08$.}
 \label{Fig4}
\end{figure}

\begin{figure}
\centering
\includegraphics[width=95mm]{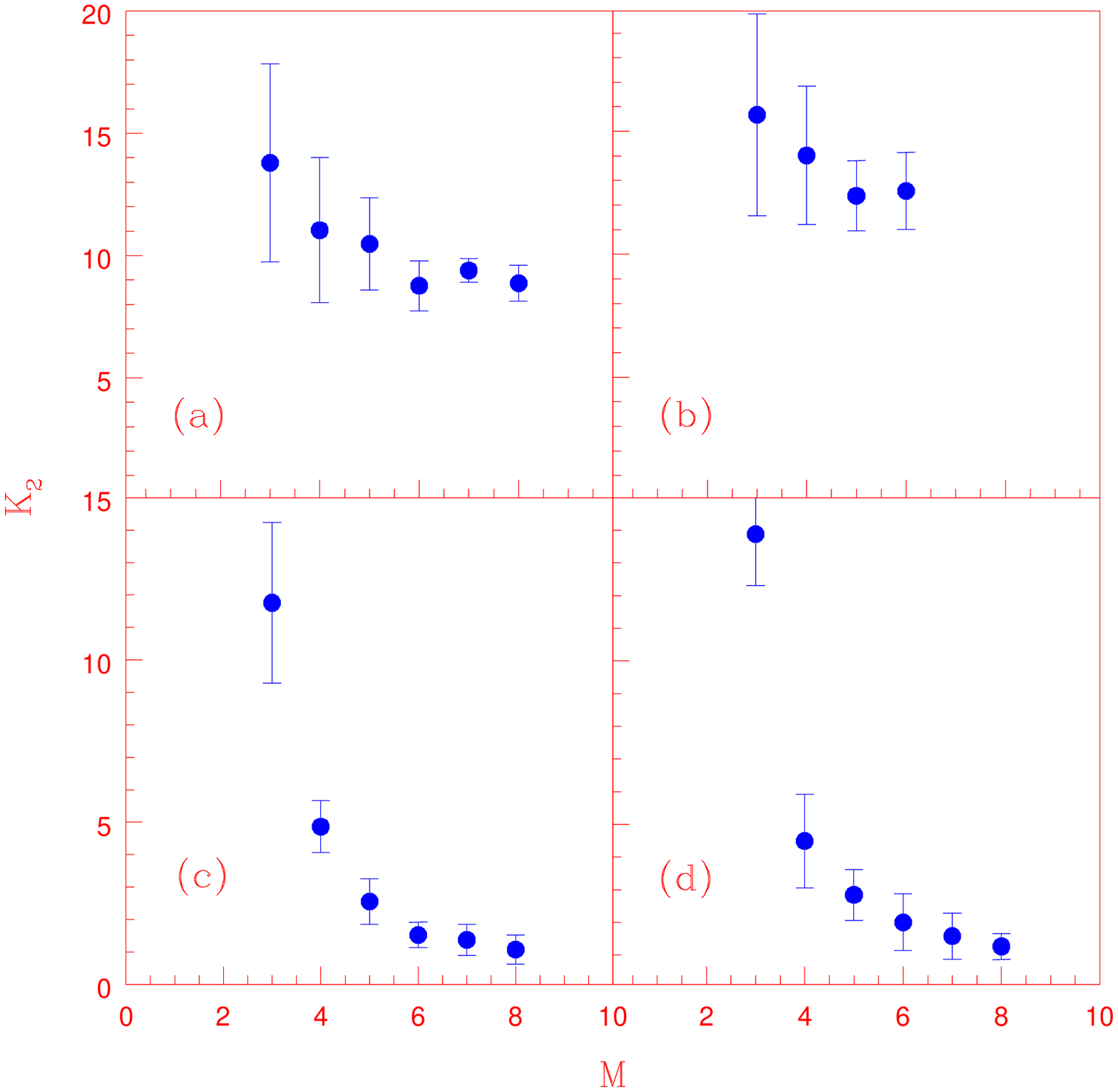}
\caption{Variation of $K_2$ with $M$ for data sets obtained by adding different 
amounts of white and colored noise to Rossler attractor data. The upper panel shows the 
result of addition of $(a) 50\%$ and $(b) 100\%$ white noise, while the lower panel shows 
the results with $(a) 50\%$ and $(b) 100\%$ colored noise with $s=2.0$. It is clear that 
while $K_2$ increases with addition of white noise, $K_2 \rightarrow 0$ as the percentage 
of colored noise increases.}
 \label{Fig5}
\end{figure}

Now the real world data is often contaminated with noise and the question that arises 
naturally is how much amount of noise can suppress the nonlinear component that may be present 
in the time series. In order to study the effect of noise on $D_2$ using our scheme, we 
generate two data sets by adding $50\%$ of white and coloured noise 
(with $s=2.0$) to the time series from the Rossler attractor. The result of applying 
our scheme on these data and their surrogates is shown in Fig.3. It is found that 
when white noise is added to the system, $D_2$ of the data increases and for a contamination 
level of $> 20\%$, it is difficult to distinguish between the data and the surrogates. But for 
colored noise contamination, the data is distinguishable from the surrogates for an added 
noise level of upto $50\%$. Note that $50\%$ noise here means that the noise amplitude is 
half of that of data. The above results indicate that from a $D_2$ analysis, it is difficult 
to distinguish even moderate amount of noise contamination in a chaotic data. A better 
quantitative measure in such a situation is $K_2$ discussed in the next section.

In order to get a quantification of the differences in the discriminating measure between 
the data and the surrogates, we use the normalised mean sigma deviation ({\it nmsd}), 
proposed by us recently \citep{kph1}. For $D_2$, this is computed using the expression 
\begin{equation}
\label{e.5}
nmsd^2   = \frac{1}{M_{max} -1} \sum_{M = 2}^{M_{max}} \Big {(}\frac{D_2 (M) - < D_2^{surr} (M) >}{\sigma^{surr}_{SD} (M) }\Big {)}^2 
\end{equation}
where $M_{max}$ is the maximum embedding dimension for which the analysis
is undertaken, $<D_2^{surr} (M)>$ is the average of $D_2^{surr} (M)$ and $\sigma^{surr}_{SD} (M)$ 
is the standard deviation of $D_2^{surr} (M)$. We have earlier shown that a value of {\it nmsd} 
$< 3.0$ implies either white or colored noise domination in the data and the null hypothesis 
cannot be rejected \citep[see for e.g.][]{kph1}.
It is found that for the Rossler attractor data shown in Fig.1, the {\it nmsd} = 36.1 
and for the two pure colored noise in Fig.2, the {\it nmsd} = 0.68 and 2.0 for 
s = 1.5 and 2.0 respectively. For data contaminated by noise in Fig.3, the 
values are 1.8 for white noise and 3.5 for colored noise.

\begin{figure}
\centering
\includegraphics[width=95mm]{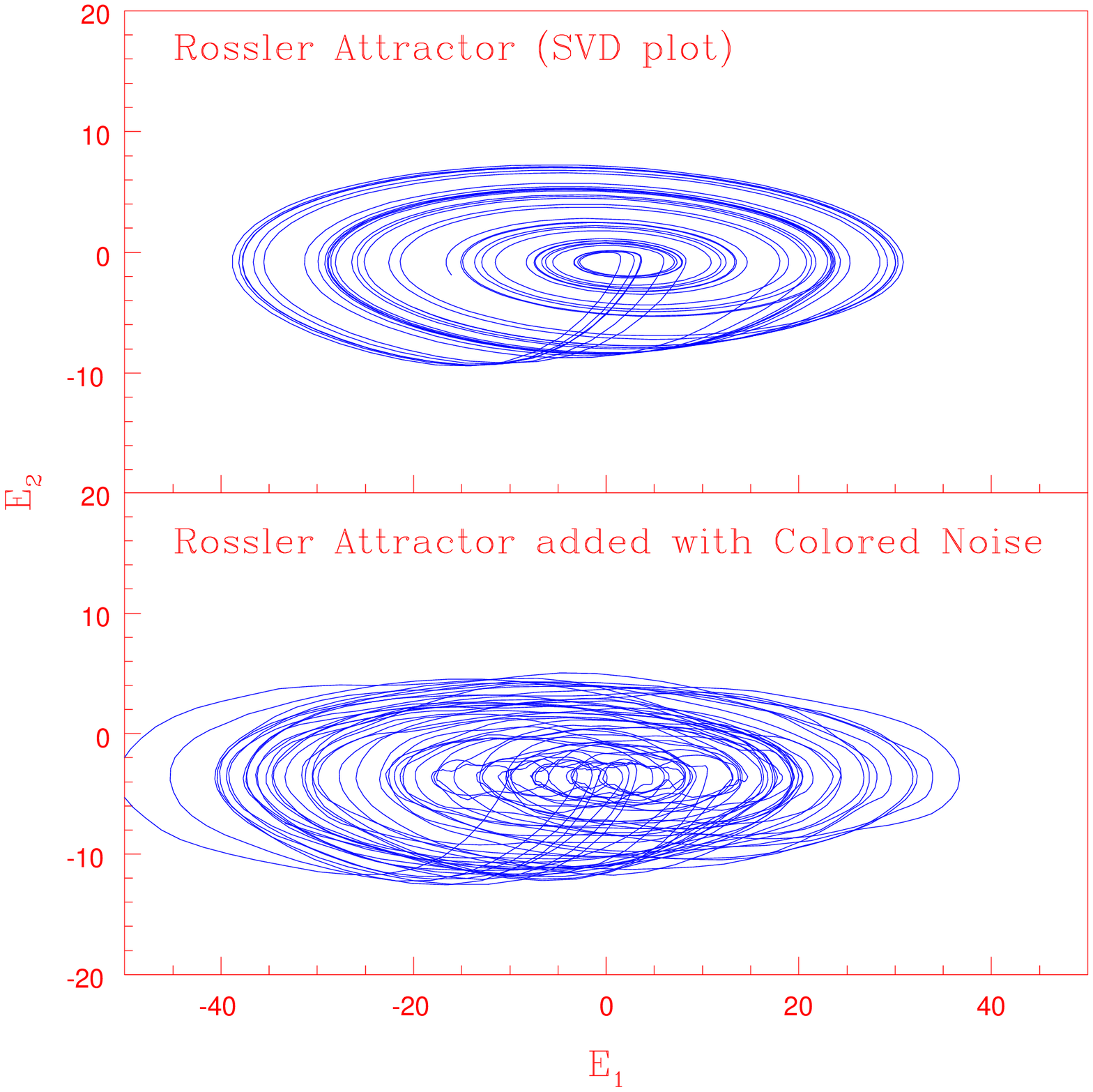}
\caption{The SVD plot of the pure Rossler attractor (upper panel) along with 
a plot of the attractor obtained by adding $50\%$ colored noise $(s=2.0)$ to the Rossler 
attractor time series.}
 \label{Fig6}
\end{figure}

\subsection{Correlation entropy}
The use of $K_2$ has been limited compared to $D_2$ for the analysis of time series data. But in 
cases where time series involve colored noise, $K_2$ is a more effective discriminating measure 
compared to $D_2$ \citep{red}. While $D_2$ is a geometric measure of the underlying chaotic 
attractor, $K_2$ is a dynamic measure representing the rate at which information needs to be 
created as the chaotic attractor evolves in time \citep{ott}. The standard method for the 
computation of $K_2$ is also the delay embedding technique. Since $K_2$ measures the rate at 
which the trajectory segments are increased as $M$ increases, it can be related to the 
correlation sum $C_M (R)$ by the expression
\begin{equation}
 \label{e.6}
 C_M (R) \propto e^{-M K_2 \Delta t}
\end{equation}
where $\Delta t$ is the time step between successive values in the time 
series. From above, a formal expression for $K_2$ can be written as, 
\begin{equation}
 \label{e.7}
 K_2 \Delta t = \lim_{R \rightarrow 0}\lim_{M \rightarrow \infty}\lim_{N \rightarrow \infty} (-{log} C_M (R)/M)
\end{equation}
Alternately, $K_2$ can also be obtained as,
\begin{equation}
 \label{e.8}
 K_2 \Delta t \equiv \lim_{R \rightarrow 0}\lim_{M \rightarrow \infty}\lim_{N \rightarrow \infty} {log} (C_M (R)/C_{M+1} (R))
\end{equation}
Our nonsubjective scheme has been extended for the computation of $K_2$ as well 
\citep{kph2} and we apply that scheme in this work.

Fig.4 shows $K_2$ for the Rossler attractor as a function of $M$ computed from the 
time series using our scheme. To show the effect of noise on $K_2$, we generate four different 
time series by adding $50\%$ and $100\%$ white as well as colored noise to the Rossler attractor 
data. The result of applying our scheme to these data sets is shown in Fig.5. It is 
clear that while the saturated $K_2$ value increases with the white noise addition, the effect of 
colored noise is quite the opposite. With the increase in colored noise, the saturated value of 
$K_2 \rightarrow 0$.

\begin{figure}
\centering
\includegraphics[width=95mm]{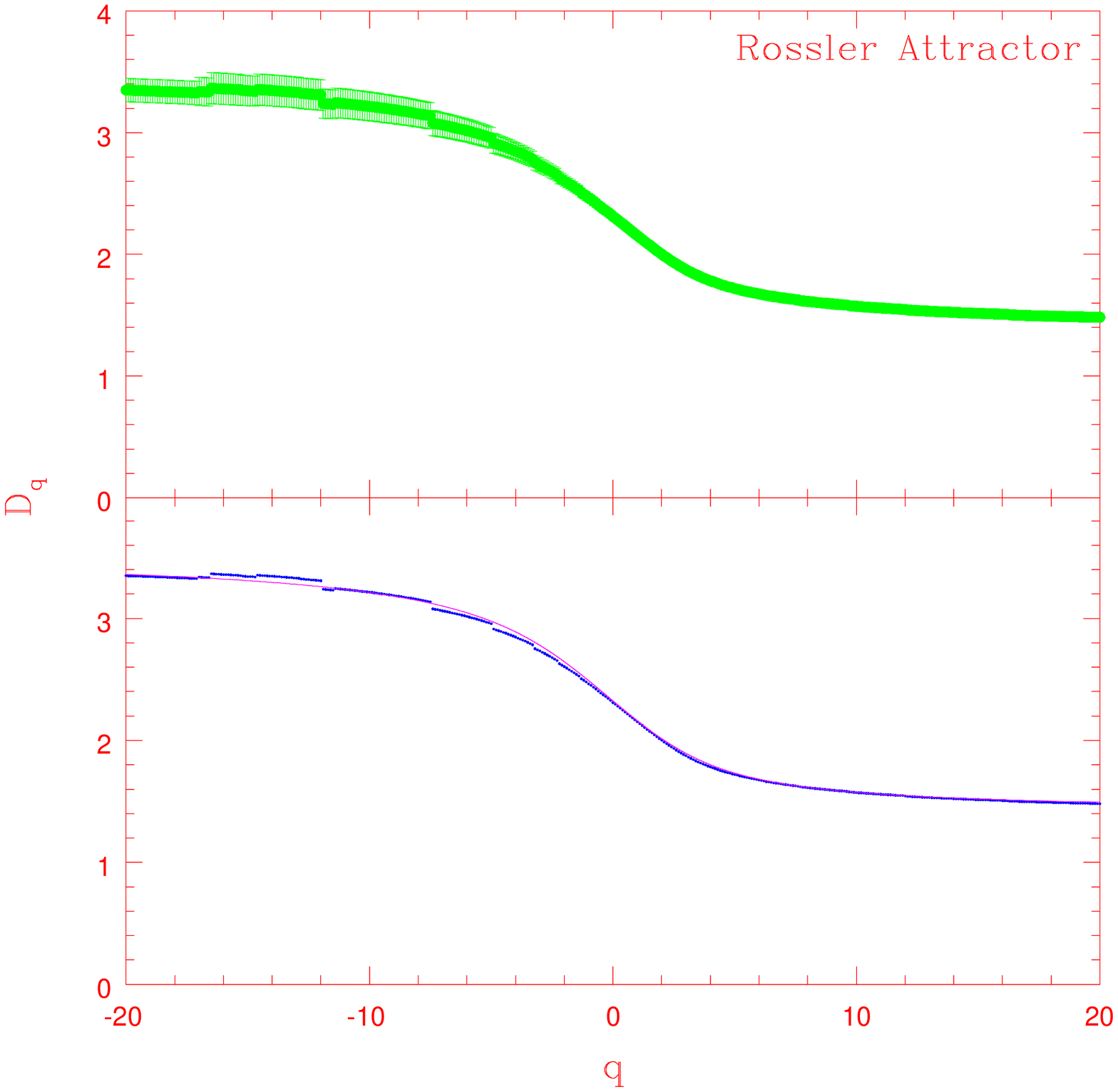}
\caption{The upper panel shows the $D_q$ spectrum of the Rossler attractor with 
error bar, computed from the time series of $10000$ data points. To show the accuracy of 
fitting, the $D_q$ values (points) are again shown in the lower panel without error bar 
along with the best fit curve (continuous line) computed using our numerical scheme.}
\label{Fig7}
\end{figure}

\begin{figure}
\centering
\includegraphics[width=95mm]{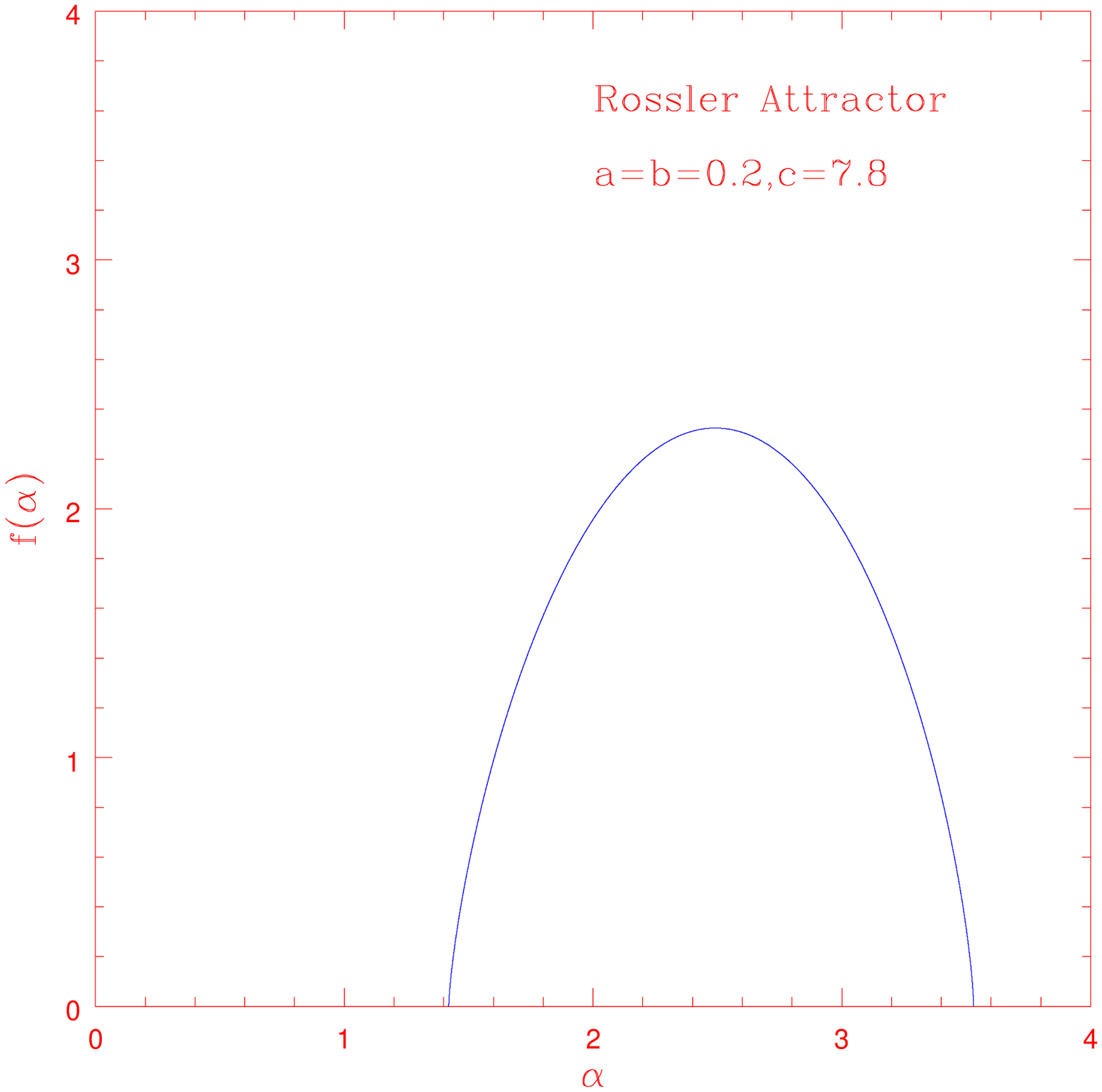}
\caption{The $f(\alpha)$ spectrum of the Rossler attractor computed from the 
best fit $D_q$ curve shown in the previous figure.}
 \label{Fig8}
\end{figure}

Our scheme can be used for surrogate analysis with $K_2$ as the discriminating measure as 
well and {\it nmsd} can be computed using a similar expression as Eq.~\ref{e.5}. We have 
applied this recently \citep{kph2} to the Rossler attractor data  with different 
percentages of white and colored noise added. For $50\%$ white noise contamination, 
{\it nmsd} with $K_2$ as the discriminating measure is found to be $4.3$, while for the 
same percentage of colored noise, the value is $2.2$. Thus, while the white noise 
contamination can be easily identified through $D_2$ analysis, the presence of colored 
noise can be better inferred by computing $K_2$. 

\subsection{Singular value decomposition}
The singular value decomposition (SVD) is another important technique used in nonlinear time 
series analysis, first proposed by \citet{bro} and for a recent review, see 
\citet{ath}. The method makes use of a trajectory matrix constructed from the experimental 
time series with the rows of the matrix constituting the state vectors in the embedding space. 
It is then diagonalised to find the dominant eigen values and eigen vectors which are used to 
represented the dynamics. The number of dominant eigen values determine the minimum number of 
dimensions required to unfold the complete dynamics and the corresponding eigen vectors give 
the projections of the reconstructed attractors. With such a SVD projection (also called 
BK projection), one can visualise the qualitative nature of the reconstructed attractors. 
Here we use the standard TISEAN algorithm \citep{heg1} for the computation of BK projections. 

\begin{figure}
\centering
\includegraphics[width=105mm]{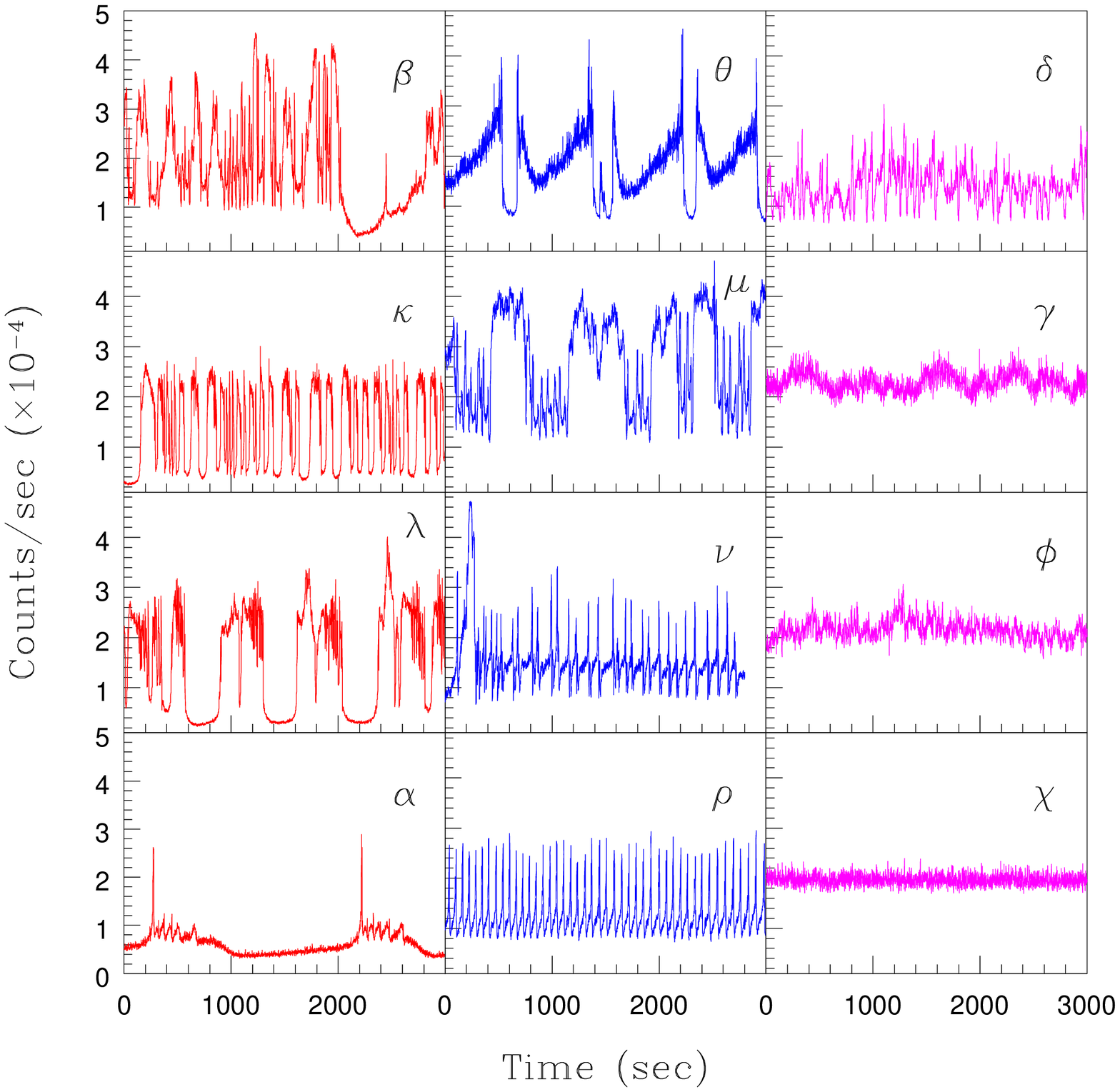}
\caption{Light curves from the twelve temporal states of the black hole system 
GRS1915+105. Only a part of the generated light curve is shown for clarity.}
 \label{Fig9}
\end{figure}

For example, the SVD projection for the Rossler attractor is shown in Fig.6 
(upper panel). To show the effect of colored noise, on the SVD projection, we also show in 
Fig.6 (lower panel) the BK projection for the Rossler attractor added with $50\%$ 
colored noise with spectral index $s=2.0$. It is evident that even such large amount of 
colored noise does not destroy the attractor completely. 

\begin{figure}
\centering
\includegraphics[width=95mm]{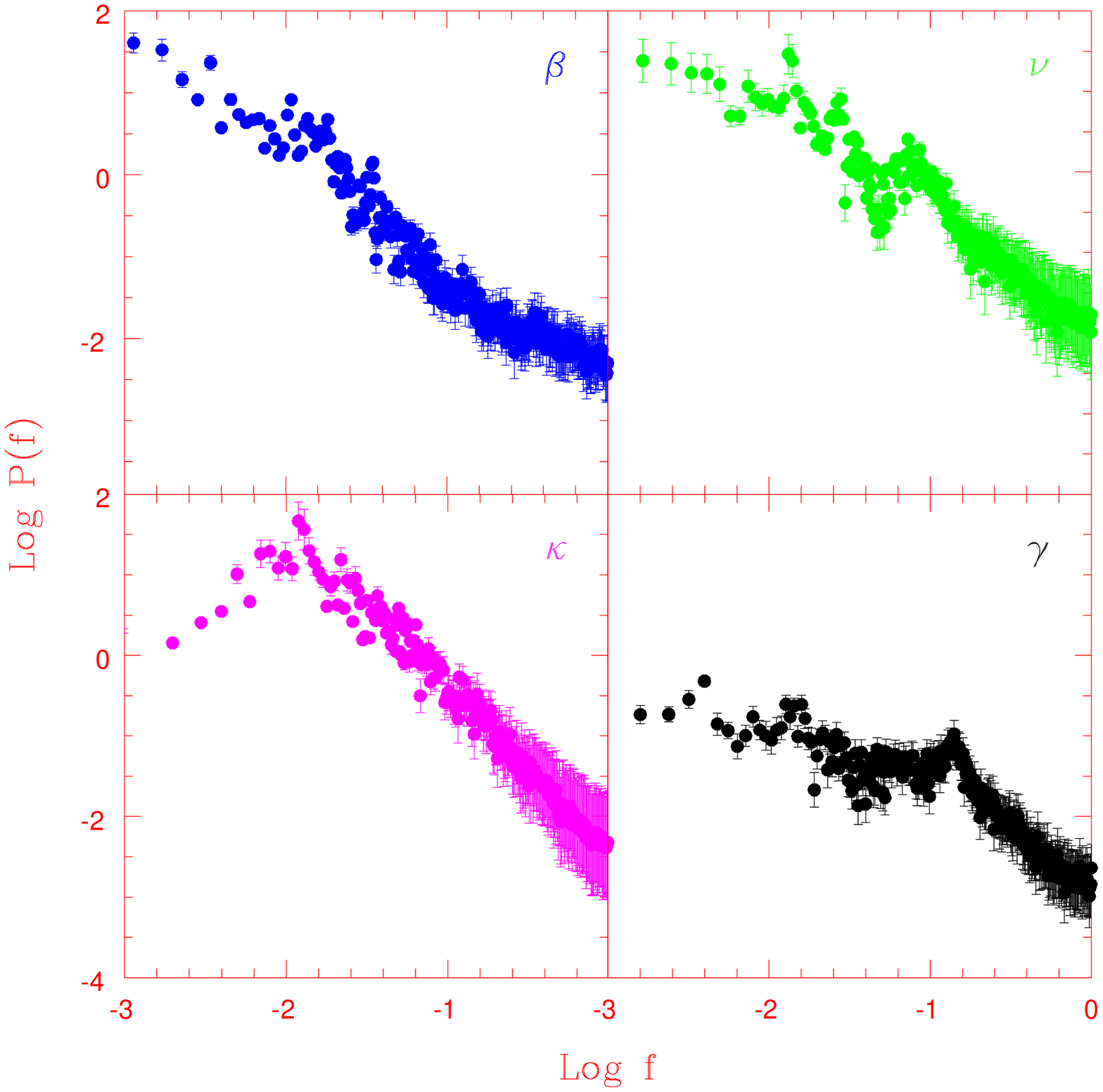}
\caption{The power spectrum for the X-ray light curves from GRS 1915+105 in 
four representative states.}
 \label{Fig10}
\end{figure}

\subsection{Multifractal spectrum}
The interest of the multifractal formalism in connection with dynamical systems 
rests on the fact that it provides us with a very efficient method to 
determine the existence of {\it strange attractors} and allows a 
statistical description of these sets. A strange or chaotic attractor normally 
possess a multifractal structure as a result of the stretching and folding of the 
trajectories in different directions in phase space. Hence they can be characterised 
by a spectrum of dimensions $D(q)$ \citep{hen}, where the index $q$ can vary from 
$-\infty$ to $+\infty$. The clustered regions of the attractor are characterised by 
$D(q)$ values with $q > 0$ and rarefied regions by $D(q)$ with $q < 0$, with $D(0)$ 
giving the simple fractal dimension of the set.

A more convenient method to represent the global scaling properties of the 
attractor is by using a spectrum of singularities characterising the probability 
measure on the strange attractor. For this, one considers the number 
$N_{\alpha}(\epsilon)$ of boxes (with edge length $\epsilon$) required to cover the 
attractor within a small range of $\alpha$ given by 
$\alpha \in [\alpha, \alpha + d\alpha]$. The parameter $\alpha$ is a continuous 
variable characterising the local scaling properties of the fractal set. Its 
meaning is that of a local scaling exponent:
\begin{equation}
 \label{e.9}
 \alpha_i =  \lim_{\epsilon \rightarrow 0} {log} p_{i} (\epsilon)/ {log} \epsilon    
\end{equation}
with $p_{i} (\epsilon)$ representing the probability measure. Thus $\alpha$ measures 
how fast the number of points within a box decreases as $\epsilon$ is reduced. It 
therefore measures the strength of a singularity for $\epsilon \rightarrow 0$. 
Assuming that a scaling hypothesis holds, one can write
\begin{equation}
 \label{e.10}
 N_{\alpha} (\epsilon) \propto \epsilon ^{-f(\alpha)}    
\end{equation}
where the exponent $f(\alpha)$ represents the fractal dimension of subsets of 
strength $\alpha$. The graph of $f(\alpha)$ as a function of $\alpha$ is called 
the $f(\alpha)$ spectrum which characterises the global scaling properties of the 
fractal set as a function of the local scaling exponents $\alpha$. The 
transformation from $D(q)$ to $f(\alpha)$ can be shown to be a Legendre 
transformation, for details, see \citet{hal} and \citet{atm2}. 

\begin{figure}
\centering
\includegraphics[width=95mm]{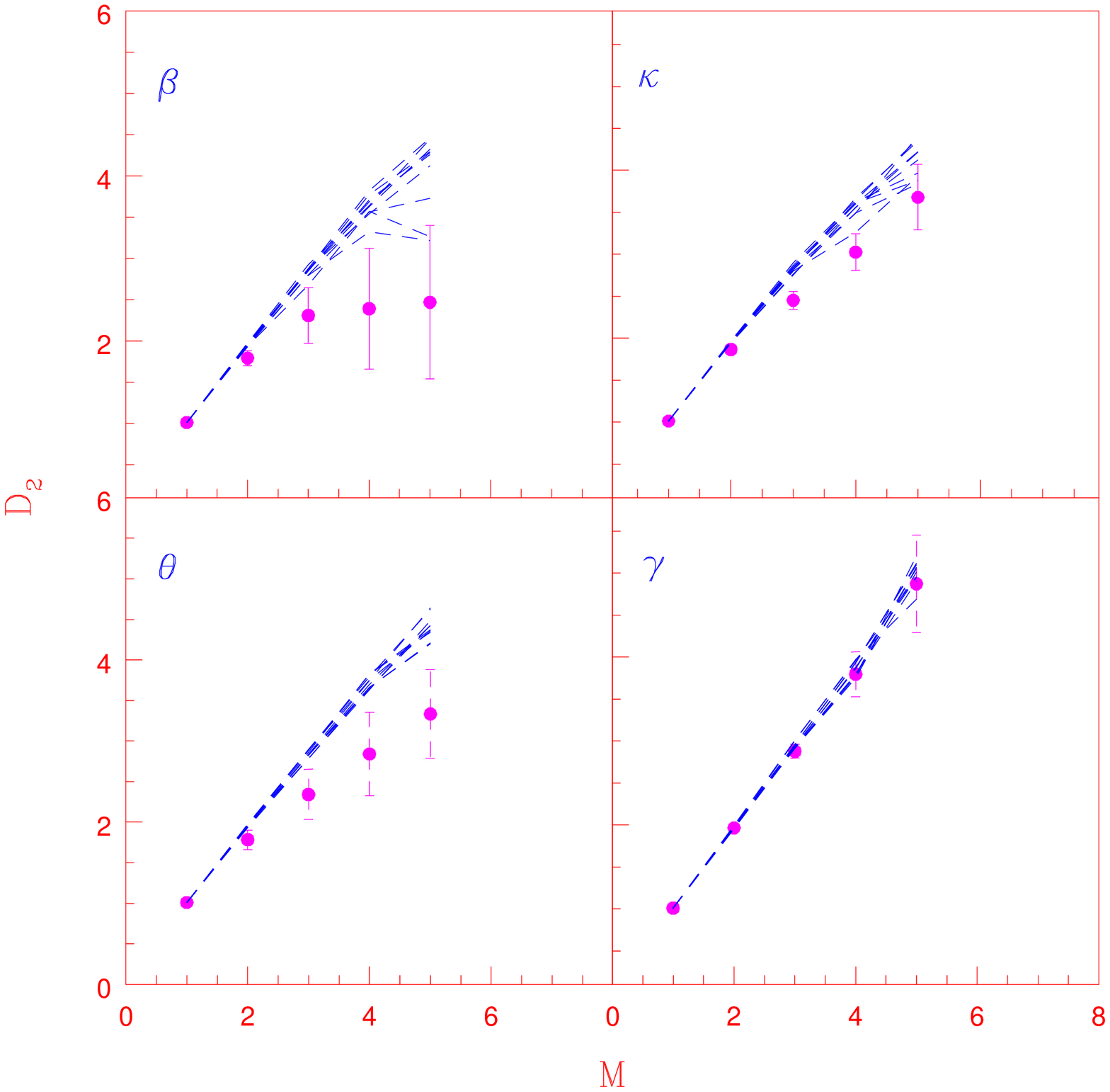}
\caption{Surrogate analysis with $D_2$ as a discriminating measure for the 
light curves from four states of GRS1915+105. Note that only the $\gamma$ state is 
consistent with noise.}
 \label{Fig11}
\end{figure}

\begin{figure}
\centering
\includegraphics[width=95mm]{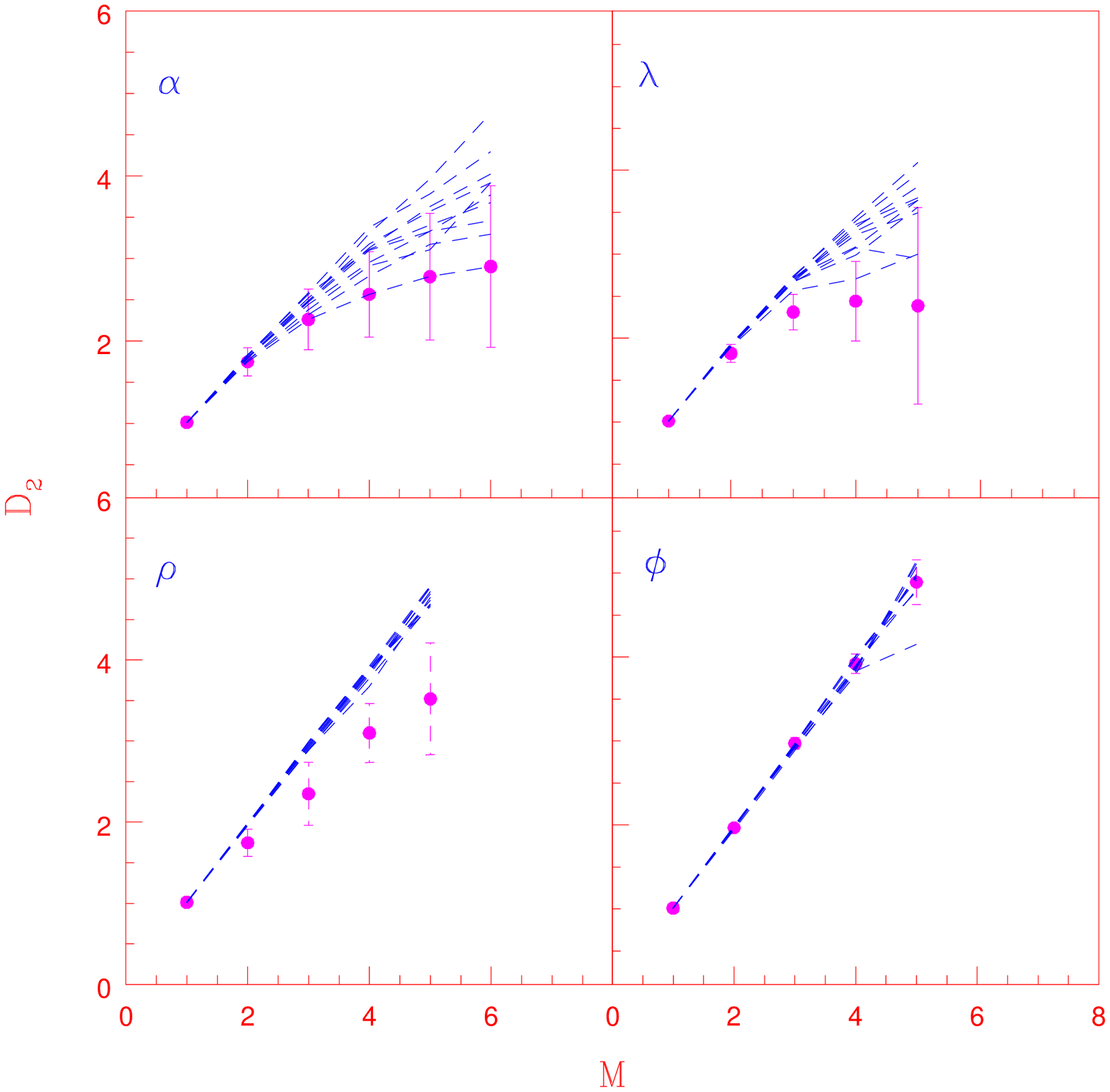}
\caption{Same as the previous figure, but with four other states. Again, 
only one state - $\phi$ - shows random behavoir.}
 \label{Fig12}
\end{figure}

To compute the $f(\alpha)$ spectrum from a time series, we first consider the 
generalised correlation sum given by
\begin{equation}
   \label{e.11}
   C_{M}^{q}(R) = [{1\over N_{c}} \sum_{i}^{N_{c}}({1\over N_v} \sum_{j=1,j 
\neq i}^{N_v} H(R-|\vec{x_{i}}-\vec{x_{j}}|))^{q-1}]^{1/(q-1)}
\end{equation}
where the Heaviside function $H$ counts how many pairs of points at 
$(\vec{x_{i}},\vec{x_{j}})$ are situated within a distance $R$.  The spectrum of 
dimensions are then determined by the relation 
\begin{equation}
    \label{e.12}
    D(q) \equiv \lim_{R \rightarrow 0} d ({log} C_{M}^{q} (R))/d ({log} (R))  
\end{equation}
The average value of $D_q$ with error bar is then calculated from the scaling region 
by taking different values of $R$, by extending the numerical procedure discussed 
above for computing $D_2$. 

In order to determine the $f(\alpha)$ spectrum, we make use of the computational scheme 
recently proposed by us and applied to several practical time series 
\citep{kph4}. 
The scheme uses an analytical fit for the $f(\alpha)$ function (involving a set of 
independant parameters) and calculates the corresponding $D_q$ curve using the Legendre 
transformation equations. This curve is then fitted to the spectrum of $D_q$ values 
computed from the time series. The best fit curve is found by changing the parameters of 
the $f(\alpha)$ fit, which is then used to compute the final $f(\alpha)$ spectrum. The 
algorithmic details of the scheme are presented elsewhere 
\citep{kph4}. The multifractal 
approach has recently been employed in the analysis of several practical time series, 
an example being the temporal variations of the geomagnetic field 
\citep{hon}. 

\begin{figure}
\centering
\includegraphics[width=95mm]{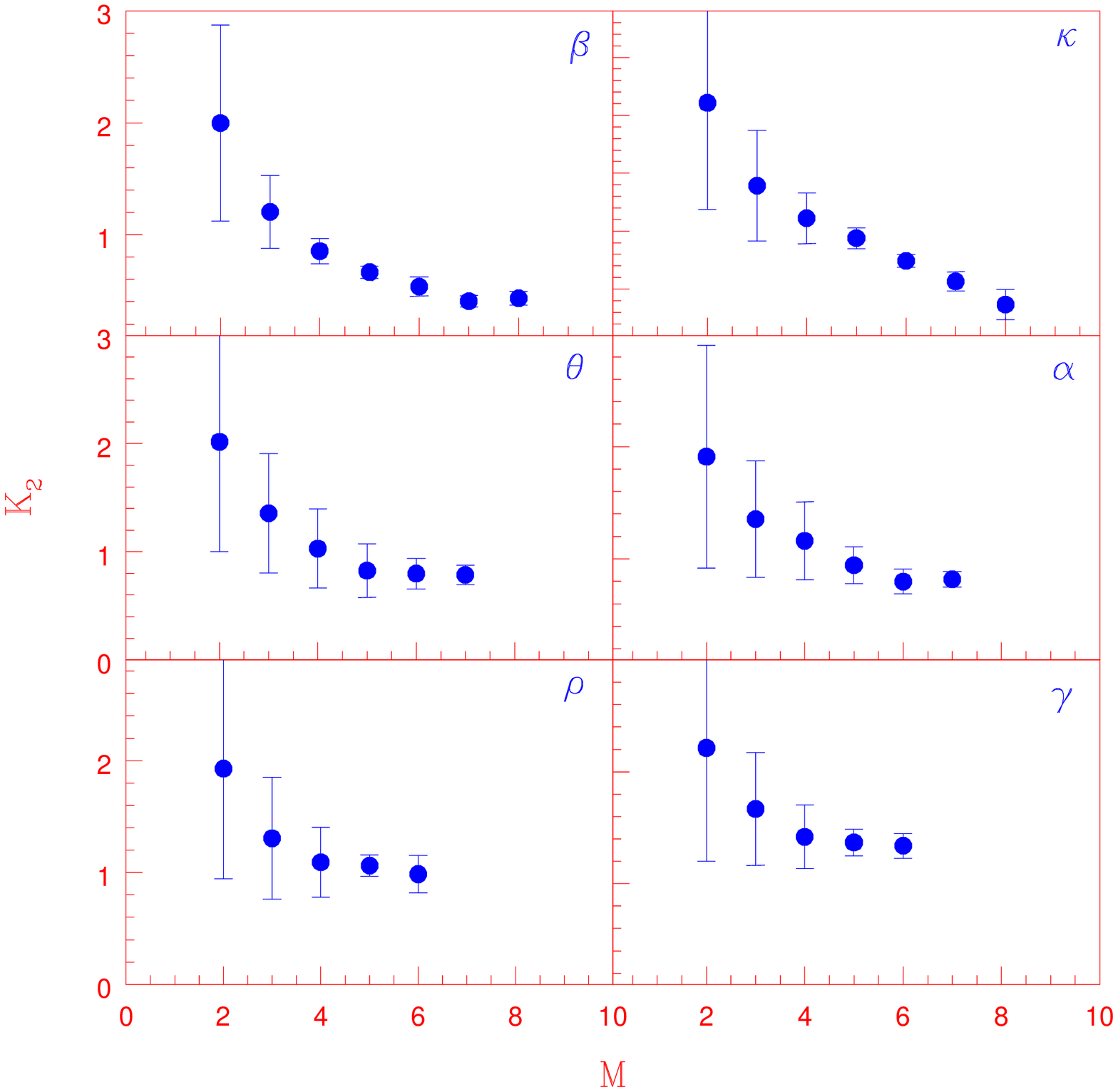}
\caption{Variation of $K_2$ as a function of $M$ for the light curves from 
six different states of GRS1915+105. While the $K_2$ values of four states converge much 
like a low dimensional chaotic system, $K_2$ for the $\kappa$ state continue to decrease 
as $M$ increases indicating colored noise contamination. Though $K_2$ for $\gamma$ state 
converge, its value is much higher and close to that of the white noise.}
 \label{Fig13}
\end{figure}

\begin{figure}
\centering
\includegraphics[width=95mm]{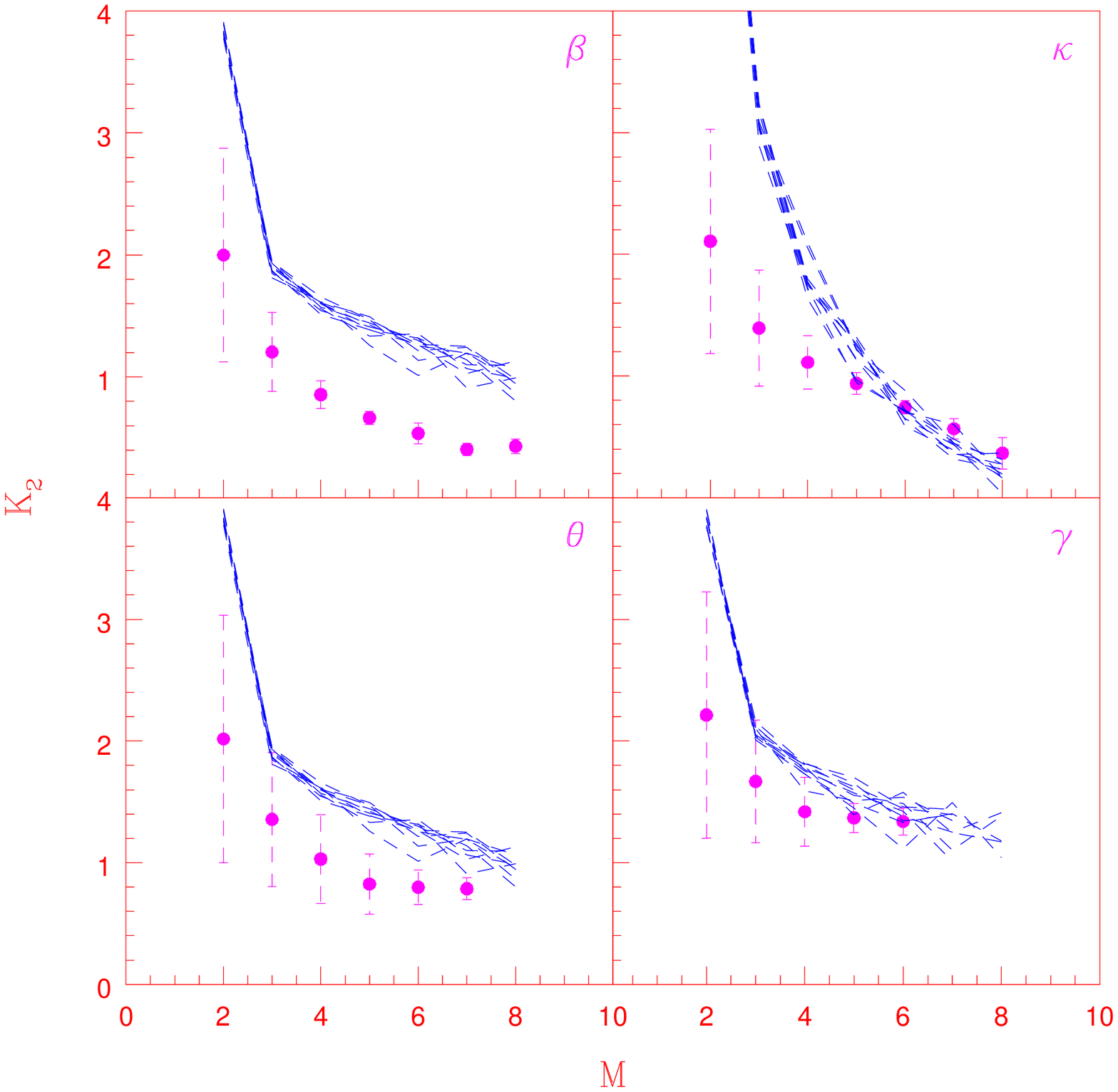}
\caption{Surrogate analysis of the light curves from four states of the 
black hole system with $K_2$ as the discriminating measure. Note that while data and 
the surrogates can be distinguished for $\beta$ and $\theta$, $\kappa$ and $\gamma$ 
behaves much like colored noise and white noise respectively.}
 \label{Fig14}
\end{figure}

To illustrate our scheme, it is first used to compute the $f(\alpha)$ spectrum of the 
Rossler attractor. The spectrum of generalised dimensions $D_q$ is computed from the time series 
taking the embedding dimension $M=3$. 
Attempting to compute the $f(\alpha)$ spectrum directly from the $D_q$ 
values  leads to an incomplete $f(\alpha)$ spectrum. This is mainly due to the fact that the 
errors in the calculation of $D_q$ makes the Legendre transformation numerically impractical 
because of the reversal of slopes. Hence our scheme uses a different procedure. 
The $f(\alpha)$ function is a single valued function defined between the
limits of $\alpha_{min}$ and $\alpha_{max}$. 
Since the derivative $f^\prime (\alpha) = d f(\alpha)/d\alpha = q$
is also single valued, it follows that $f (\alpha)$ has a single extremum (i.e. a maximum).
Moreover, $f(\alpha_{min}) = f (\alpha_{max}) = 0$ and $f^\prime (\alpha_{min})$ and
$f^\prime (\alpha_{max})$ tend to $ \infty$ and $ - \infty$ respectively. A simple function
which can satisfy all the above necessary conditions is,
\begin{equation}
   \label{e.13}
   f(\alpha) = A(\alpha - \alpha_{min})^{\gamma_1}(\alpha_{max} - \alpha)^{\gamma_2}
\end{equation} 
where $A$, $\gamma_1$, $\gamma_2$, $\alpha_{min}$ and $\alpha_{max}$ are a 
set of parameters characterising a particular $f(\alpha)$ curve. 
The $D_q$ curve can be computed from this $f(\alpha)$ fit using the inverse Legendre transformation 
equations for a given set of parameters. It is then fitted to the $D_q$ spectrum computed 
from the time series. The statistically best fit $D_q$ curve is found by adjusting the parameters 
of the $f(\alpha)$ function, which is then used to compute the final $f(\alpha)$ spectrum. 
The $D_q$ spectrum and its best fit curve for the Rossler attractor are shown in 
Fig.7 and the $f(\alpha)$ spectrum computed from the best fit curve is 
shown in Fig.8. Having discussed the various measures and schemes for computing them, 
we now turn to the analysis of the black hole system GRS1915+105.

\begin{figure}
\centering
\includegraphics[width=95mm]{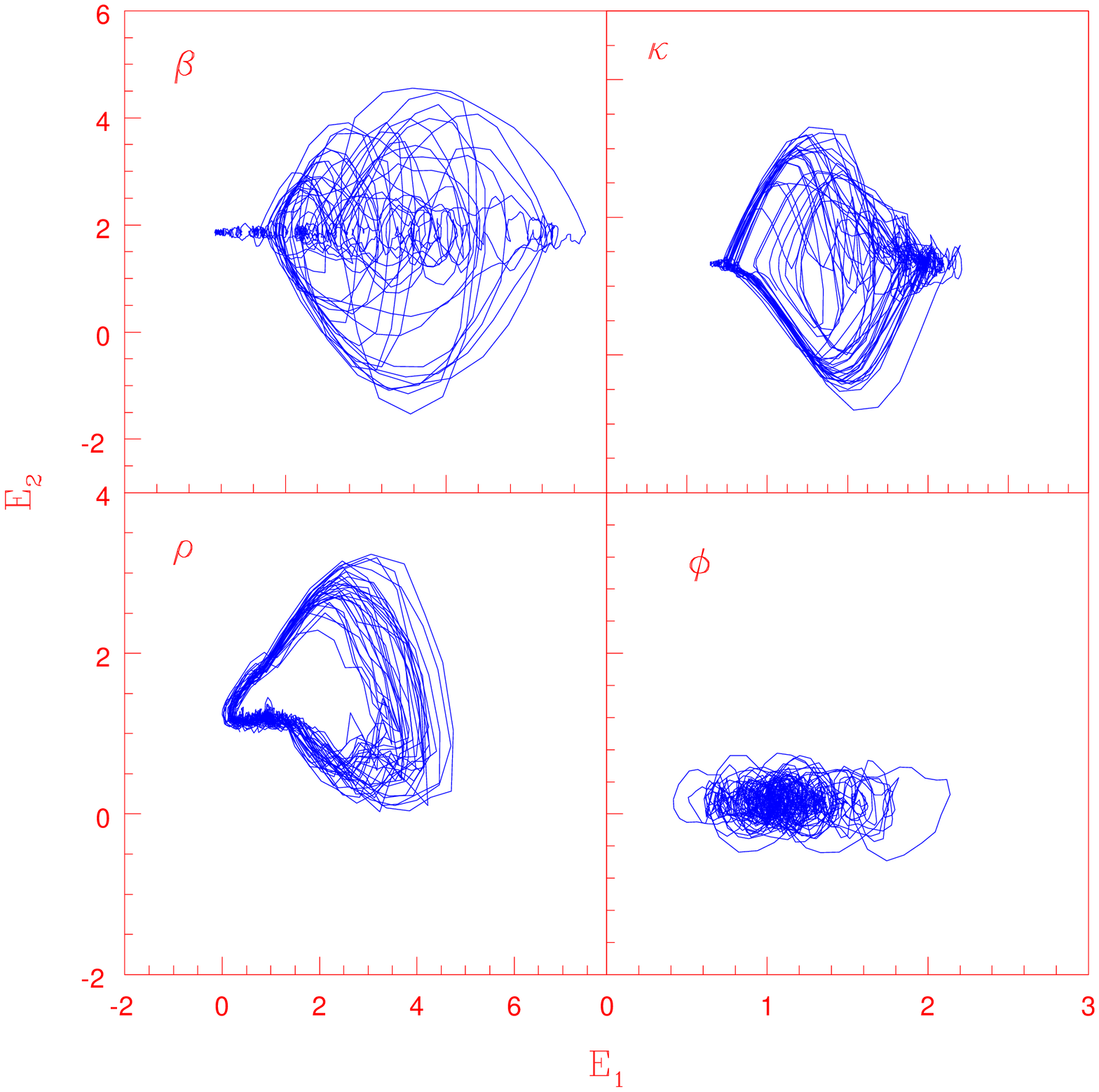}
\caption{The plot of the attractors underlying four states of the black hole 
system reconstructed via SVD analysis. Except the $\phi$ state, which behaves similar to a 
white noise, all the others indicate the presence of underlying attractors, the most 
interesting being the $\rho$ state.}
 \label{Fig15}
\end{figure}

\section{Analysis of the Black Hole System GRS1915+105}
\label{sect:anal}
In this section, we apply all the techniques
discussed above to analyse the X-ray light  curves from the black hole
binary GRS1915+105. The temporal properties of the system have  been
classified into 12 different spectroscopic classes by \citet{bel}
based on the RXTE data. Here we have chosen  representative data sets
for each class. The light-curve for an observation was obtained from
the standard products\footnote{http://heasarc.gsfc.nasa.gov/docs/xte/recipes/stdprod\_guide.html}
which provide  a 0.125 secs time resolution summed over all energy channels. 
Standard product
light curves have been generated using a pipeline which considers standard
filtering criteria and use data from the instruments that were reliably on.
While standard products may not have optimal spectral response matrix or
background model, they are more than adequate for light curve analysis,
especially for bright sources, like GRS 1915+105, where the background is
not important.

The analysis requires continuous data without gaps. For each class, we have extracted two 
sets of continuous segments for the analysis. The light curves have been generated 
after rebinning to a time resolution of 0.5 secs resulting in $\sim 5200$ to 
$6400$ continuous data points for each segment. Light curves with finer time 
resolutions are more Poisson noise dominated, while larger binning gives less data points. 
Table 1 gives the observation ID, class, number of data points etc. of all the light curves 
used for the analysis. In the last column, we also provide the temporal behavior of the 
light curve as resulted from our analysis. More details regarding the data, such as, 
average count, expected Poisson noise variation, etc. are given elsewhere \citep{mis2}.

\begin{table}[h!!!]

\small
\centering

\begin{minipage}[]{120mm}

\caption[]{ Details of the light curves from GRS 1915+105, in all the 12 spectral 
classes, used for the analysis. For each class, light curves from two Observation IDs 
have been analysed, as indicated. The second column gives the number of continuous 
data points after rebinning. The last column indicates the temporal 
behavior of the class as obtained from our analysis.}\end{minipage}

\small
 \begin{tabular}{cccc}
  \hline
  \hline\noalign{\smallskip}
{\it Obs. ID}  &  {\it No. of Data Points} & {\it Class} & {\it Temporal Bahavior}  \\ 
\hline\noalign{\smallskip}
10408-01-10-00  & 6146 &           &                                    \\
20402-01-46-00  & 5504 &  $\beta$  &  Deterministic Nonlinear           \\
&&                                                                      \\
20402-01-45-00  & 6156 &           &                                     \\
10408-01-15-00  & 5764 &  $\theta$ &  Deterministic Nonlinear                   \\
&&                                                                             \\
20402-01-03-00  & 6244 &           &                                           \\
20402-01-31-00  & 5876 &  $\rho$   &  Deterministic Nonlinear                  \\
&&                                                                             \\
10408-01-40-00  & 6024 &           &                                           \\
10408-01-41-00  & 5312 &  $\nu$    &  Deterministic Nonlinear                  \\
&&                                                                             \\
20187-02-01-00  & 6010 &           &                                            \\
20402-01-22-00  & 5220 &  $\alpha$ &  Deterministic Nonlinear                   \\
&&                                                                               \\
20402-01-33-00  & 6240 &           &                                             \\
20402-01-35-00  & 6244 &  $\kappa$ &  Deterministic Nonlinear + Colored Noise    \\
&&                                                                               \\
20402-01-37-00  & 6080 &           &                                              \\
20402-01-36-00  & 5648 &  $\lambda$ & Deterministic Nonlinear + Colored Noise     \\
&&                                                                                 \\
10408-01-08-00  & 5688 &           &                                              \\
10408-01-34-00  & 5756 &  $\mu$    &  Deterministic Nonlinear + Colored Noise     \\
&&                                                      \\
10408-01-17-00  & 6010 &           &                     \\
20402-01-41-00  & 5466 &  $\delta$ &  White Noise         \\
&&                                                       \\
20402-01-56-00  & 6324 &           &                     \\
20402-01-39-00  & 6180 &  $\gamma$ &  White Noise         \\
&&                                                       \\
10408-01-12-00  & 6286 &           &                     \\
10408-01-09-00  & 5580 &  $\phi$   &  White Noise         \\
&&                                                        \\
10408-01-22-00  & 6022 &           &                     \\
20402-01-04-00  & 5382 &  $\chi$   &  White Noise         \\ 
  \noalign{\smallskip}\hline
\hline
\end{tabular}
\end{table}

Fig.9 shows all the 12 light curves used for the analysis, which are labelled by 12 
different symbols representing the 12 temporal states of the black hole system. 
We show only one set of light curve since the second set looks identical 
for all the states. The system  
appears to flip from one state to another randomly in time. The classification of 
\citet{bel} is based on a detailed analysis of all the light curves from RXTE data 
using various linear tools. But it is difficult to differentiate the subtle 
temporal features between the light curves with the help of the linear tools, 
such as, the power spectrum. For example, in Fig.10, we show the power spectrum 
for four representative states, whose temporal properties are different as per 
our analysis (see Table 1). While $\beta$ and $\nu$ are candidates for deterministic nonlinear 
behavior, $\kappa$ is possibly a mixture of nonlinearity and colored noise and 
$\gamma$ is purely stochastic. But these distinctions are barely evident from the 
power spectral variations, though the $\gamma$ state appears more like a white 
noise, in agreement with our results.  
This, once again, emphasizes the importance of methods 
based on nonlinear time series analysis for a better understanding of the temporal 
properties of the light curves.

Recently, we applied surrogate 
analysis on all these light curves and showed that more than half of these 12 states deviated from 
a purely stochastic behavior \citep{mis1}. 
Here we combine the results of computations of $D_2$, 
$K_2$ and SVD analysis to get a better understanding regarding the nature of these light curves. 
We have done the surrogate analysis with $D_2$ and $K_2$ and the SVD analysis separately for 
the two sets of light curves. But here we only show plots from representative light 
curves from one set as the plots from the second set are similar and the results are 
qualitatively the same.  
Figs.11 and 12 show the results of surrogate analysis, with $D_2$ as 
discriminating measure, on eight different states. Of the  states shown in these figures, 
it clear that only two states - $\gamma$ and $\phi$ - show purely stochastic behavior.

\begin{figure}
\centering
\includegraphics[width=95mm]{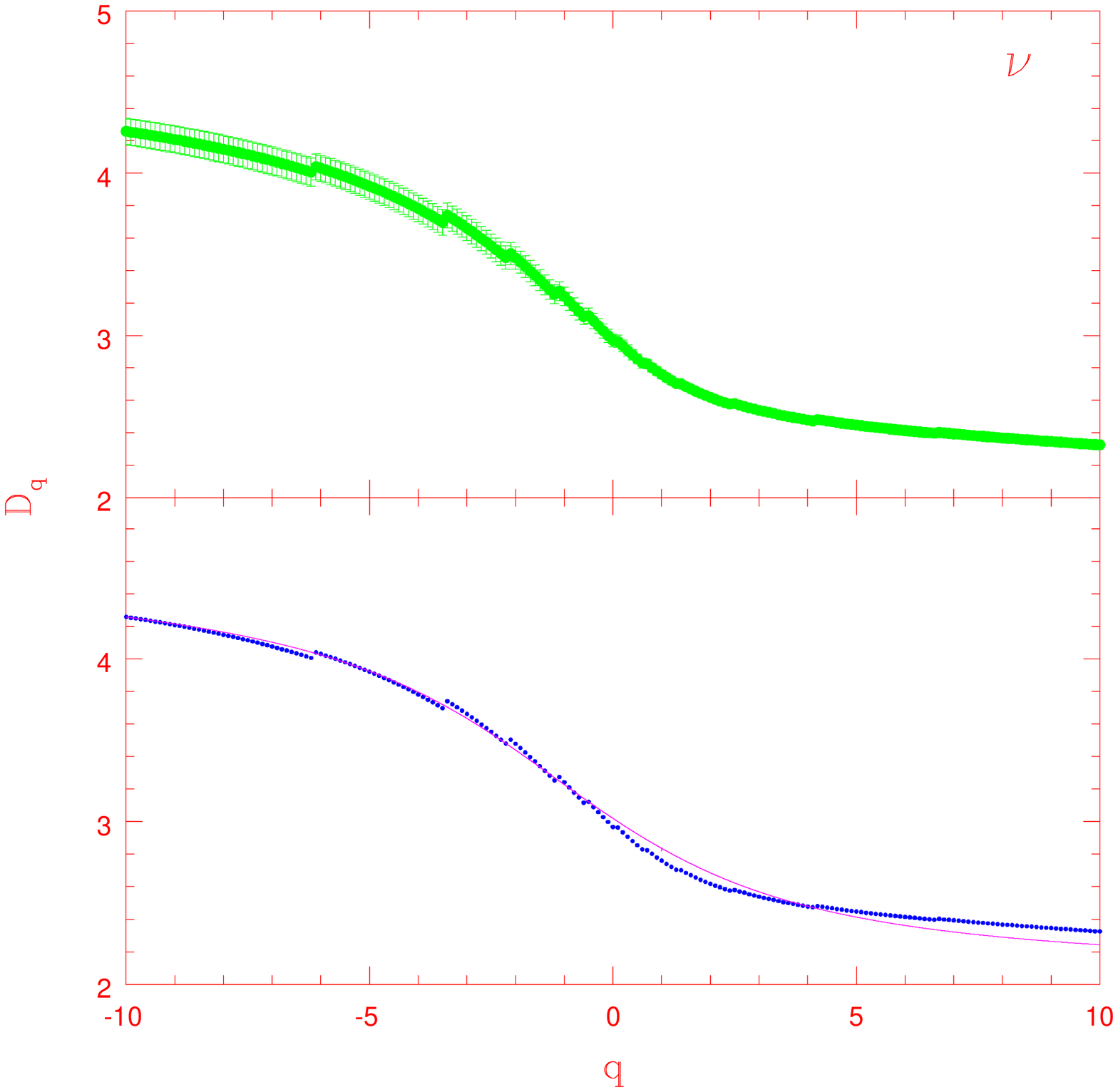}
\caption{The spectrum of generalised dimensions with error bar (upper panel) 
for the $\nu$ state corresponding to embedding dimension $M=3$. The lower panel shows the 
$D_q$ values without error bar and the best fit curve.}
 \label{Fig16}
\end{figure}

\begin{figure}
\centering
\includegraphics[width=95mm]{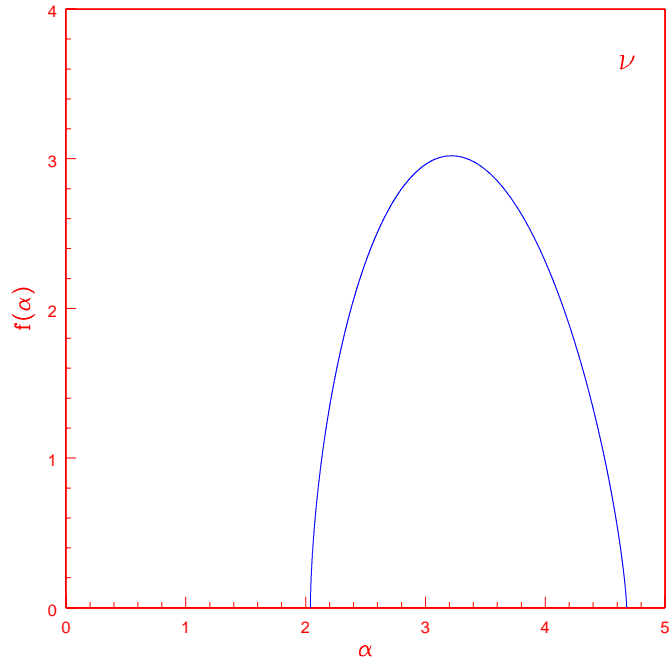}
\caption{The multifractal spectrum for the $\nu$ state computed from the 
best fit curve for $D_q$ shown in the previous figure.}
 \label{Fig17}
\end{figure}

\begin{table}

\bc

\begin{minipage}[]{120mm}

\caption[]{ The values of {\it nmsd} computed for all the 12 GRS states 
from the two sets of light curves with 
$D_2$ and $K_2$ as discriminating measures.}\end{minipage}

\small
 \begin{tabular}{ccccc}
 \hline
  \hline\noalign{\smallskip}
{\it GRS state}  &  $nmsd(D_2)$ &  $nmsd(D_2)$ &  $nmsd(K_2)$ &  $nmsd(K_2)$  \\
                 &   set 1      &   set 2      &    set 1     &    set 2     \\ 
\hline\noalign{\smallskip}
$\beta$           & 7.04        & 9.32         & 13.74       & 11.44       \\
$\theta$           & 10.63      & 8.59         & 11.20       & 10.86       \\
$\alpha$           & 8.18       & 6.71         & 8.92        & 6.89        \\
$\nu$           & 5.94          & 6.04         & 6.87        & 6.35        \\
$\rho$           & 11.25        & 10.83        & 14.28       & 12.08       \\
$\kappa$           & 4.64       & 4.77         & 3.22        & 3.54        \\
$\lambda$           & 6.66      & 6.22         & 4.57        & 4.67        \\
$\mu$           & 4.86          & 4.90         & 3.98        & 3.82        \\
$\delta$           & 2.32       & 3.13         & 1.34        & 1.68        \\
$\gamma$           & 0.88       & 1.03         & 1.83        & 1.43        \\
$\phi$           & 0.96         & 0.92         & 2.12        & 1.74        \\
$\chi$           & 0.78         & 0.73         & 1.67        & 1.38        \\ 
  \noalign{\smallskip}\hline
\hline
\end{tabular}
\ec
\end{table}

Fig.13 presents the results of computation of $K_2$ for six of the above eight states.
Since colored noise is also expected in the black hole data, surrogate analysis has been performed 
with $K_2$ as the discriminating measure on all the 12 states. The results are shown in 
Fig.14 for four of these states. While the behavior of $\beta$, $\theta$ and 
$\gamma$ are consistent with earlier analysis, the behavior of $\kappa$ suggests that it is 
contaminated by colored noise. 

To get a quantitative measure, we now compute {\it nmsd} using both $D_2$ and $K_2$ as 
discriminating measures for all the 12 states from the two sets of light curves 
and the results are shown in Table 2. 
A careful inspection of the Table reveals the following results. 
The values of {\it nmsd} in the two cases suggest that the temporal properties of 
the light curves in the two sets are almost identical. Out of the 12 states, 
four $(\delta, \gamma, \phi, \chi)$ are completely stochastic or white noise.  
Of the remaining eight 
states, three - $\kappa, \lambda$ and $\mu$ - are contaminated by colored noise and the 
rest - $\beta, \theta, \alpha, \nu$ and $\rho$ - show signatures of deterministic 
nonlinear behavior in their temporal variations. It is generally expected that all 
the states contain some amount of white noise whose percentage may vary. For example, 
in the case of the state $\alpha$, 
the saturated values of $D_2$ and $K_2$ improves significantly as the resolution time is 
increased from 0.5 sec to 1 sec. This clearly indicates the presence of Poisson white noise 
in the data. But the surrogate analysis with both $D_2$ and $K_2$ confirms that the 
null hypothesis can be rejected for the light curve in the $\alpha$ state.

We next perform a SVD analysis on all the states which clearly show the qualitative 
nature of the underlying attractors. The plot of attractors for selected states are shown in 
Fig.15. The most interesting plot is for the $\rho$ state which shows a typical 
limit cycle type attractor. Also, note that the SVD plot for $\kappa$ has a nontrivial 
appearance, even though surrogate analysis suggested the presence of colored noise. This is 
also true in the case of the other two identical states, $\lambda$ and $\mu$. Thus these 
three states could be a mixture of deterministic nonlinearity and colored noise. 
Thus, based on our results, the 12 states can be divided into 
3 broader classes from the point of view of their temporal properties. It turns out that 
some of these states which are spectroscopically different, behave identically in their 
nonlinear dynamics characteristics. This may be an indication of some common features in the 
mechanism of production of light curves from these states. The temporal behavior of 
each state as obtained from our analysis is indicated in the last column in Table 1. 
Since the behavior is identical for two different observation IDs in all cases, it 
may be concluded that the results presented here are not dependent on sample 
selection and are applicable for all the light curves classified by \citet{bel}.

\begin{figure}
\centering
\includegraphics[width=95mm]{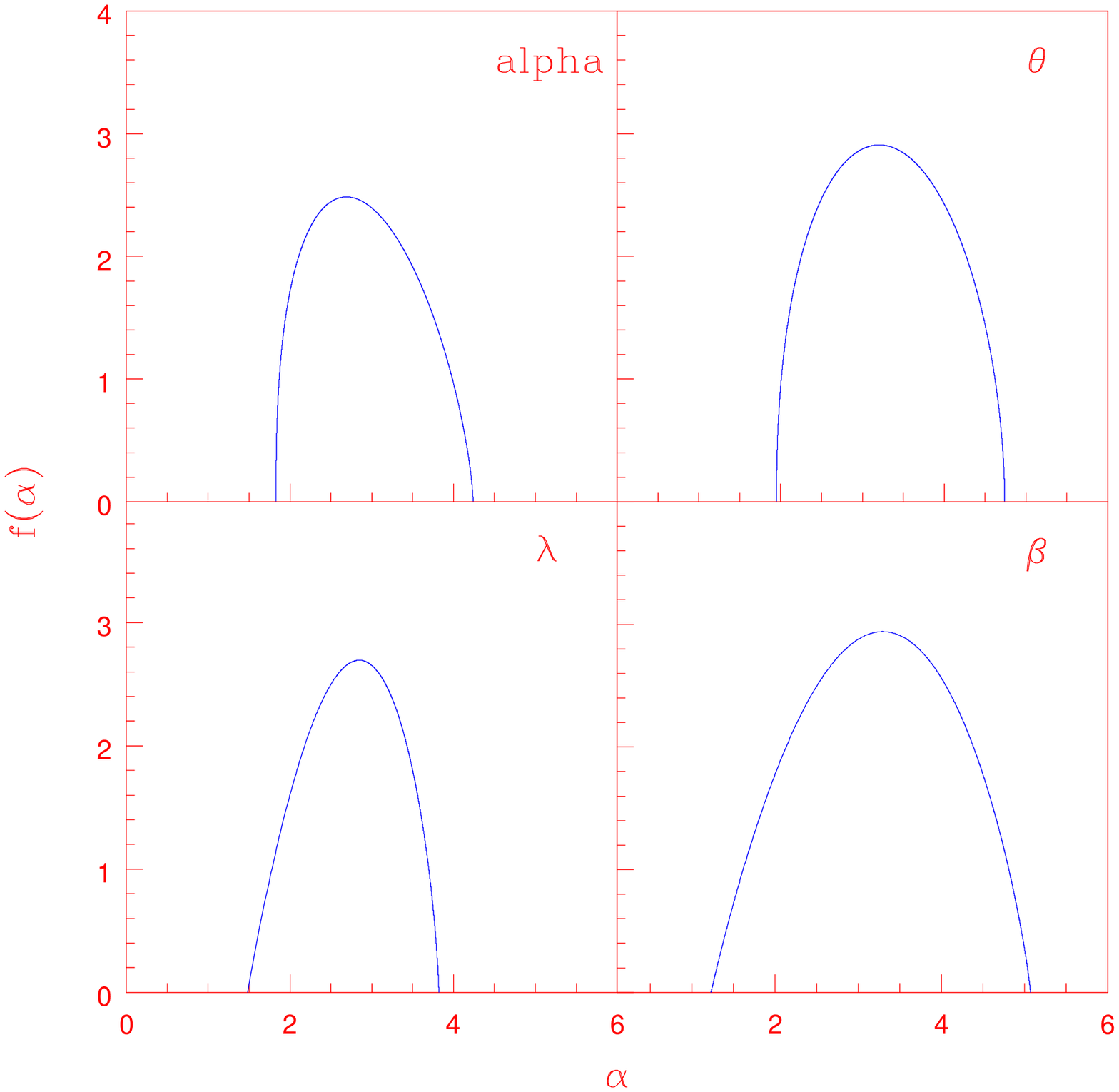}
\caption{The $f(\alpha)$ spectrum for the light curves corresponding to 
four states of the black hole system computed using our scheme with $M=3$. Note that 
GRS state has been labelled {\it alpha} in order to avoid confusion with the 
scaling index $\alpha$.}
 \label{Fig18}
\end{figure}

Finally, we show the results of multifractal analysis of all the light curves except the 
four which show purely stochastic behavior and hence the $f(\alpha)$ spectrum is 
irrelevant. Our non subjective scheme for computing $D_q$ and $f(\alpha)$ spectrum discussed in 
Sec. 2.4, provides us with a set of parameters that can be used to compare the fractal 
properties between different states as reflected in the light curves. To show the 
details of the computations, we first take a typical state. In Fig.16, we show the 
$D_q$ spectrum for the $\nu$ state  
along with the best fit curve from which the $f(\alpha)$ is  
computed as given in Fig.17. This is repeated for the other states as well and 
the results for four other states are shown in Fig.18.
The multifractal nature of the attractors is evident from the figures. Since the 
computation is done under fixed conditions prescribed by the algorithmic scheme, the 
associated parameters characterising the spectra can give a better representation for 
comparison between various states. Our results indicate that the spectra and the 
associated parameters are typically different for each state and do not show any clear 
trend among members that display strong deterministic nonlinear behavior. This can be 
inferred from Fig.18 for the case of range of scales. Thus, it turns out 
that there are subtle differences between the states belonging to the same dynamic 
class with respect to multifractal scaling as well, apart from linear spectral 
characteristics based on which the 12 states are divided.

\section{Discussion and Conclusion}
\label{sect:dicussion}
Identifying nontrivial structures in real world systems is considered to be a 
challenging task as it requires a succession of tests using various quantitative 
measures. Eventhough a large number of potential systems from various fields have been 
analysed so far, the results remain inconclusive in many cases. Here we present an 
example of a very interesting astrophysical system, which we analyse using several 
important quantifiers of low dimensional chaos.
By using the time series from a standard chaotic system - the Rossler 
attractor - we first test the computational schemes used for the analysis. 
These schemes are then applied to the light curves from the black hole system. We find that 
out of the 12 spectroscopic states, only four are purely stochastic. The remaining show 
signatures of deterministic nonlinearity, with three of them contaminated by colored 
noise. All these 8 states are found to have $D_2 < 4$ so that their complex temporal 
behavior can be approximated by 3 or 4 coupled ordinary differential equations. Based on 
our results, the 12 states can be broadly classified into three from a dynamical 
perspective: purely stochastic with 
$D_2 \rightarrow \infty$, affected by colored noise and those which are potential 
candidates for low dimensional $(D_2 < 4)$ chaotic behavior.  

It should be noted that \citet{bel} classified the light curves into 12 
states based on their count rate, variability and spectral characteristics. In other 
words, this is a classification based on linear characteristics of the light curves. 
Ours is not a classification as in the strict sense of \citet{bel}. We only show 
that some of the light curves which appear different based on their variability and 
spectral properties can be grouped together when viewed from a dynamical perspective. 

Our results could be significant in many ways. First of all, this is 
the first real evidence of a possible 
multifractal attractor in the time series of a black hole system. The fact that the light curves 
from many of the temporal states have underlying strange attractor like behavior, increases 
the possibility that the temporal variability in the time scales within these states are 
governed by some inherent nonlinear processes with a few degrees of freedom. 
In other words, the complex nonlinear partial differential equations that are known to 
govern the hydrodynamic flow can be approximated by a set of ordinary differential 
equations and hence can be more easily studied and understood. Moreover, the 
result that some of the states which are spectroscopically different,  have approximately 
the same nonlinear characteristics is interesting from certain dynamical aspects, such as 
the mechanism of production of light curves. It is well known that GRS1915+105 is a unique 
black hole system with many temporal states which vary over a wide range of time scales. 
Many of the questions regarding this variability and the exact mechanism of production of 
light curves still remain unanswered. 

Another question is regarding the structure of variability between the 12 spectroscopic 
states. It has been suggested that all the observed light curves could be interpreted 
in terms of three basic states (a hard state and two softer states) and a sequence of 
transitions between them \citep{bel}. This could, in principle, give rise to a much 
larger variety of light curves. But the system chooses only a handful of these 
sequences. This possibly suggests that the structure of time variability is not 
random, but controlled by some physical parameter which must be connected to the basic 
properties of the accretion disk. The presence of deterministic nonlinear behavior in 
the system further substantiates this idea.

Interestingly, the system appears to be in the $\chi$ state for most of the time which is 
identified as purely stochastic in our analysis. We have also analysed many samples of the 
long time average of the light curves from the system and found that they show purely 
random bahavior. Thus, one possibility is that the states other than the $\chi$ state 
may well be short time {\it flips} due to some changes taking place within the system. 
But many of these short time states pick up much less amount of noise revealing, for 
example, the underlying nonlinear character. Thus, an interesting question is whether 
the states such as $\beta$ and $\theta$ have a different underlying mechanism of 
production of light curves compared to the $\chi$ state. Or, is the excessive amount of 
white noise in that state itself is suppressing the nonlinear properties? The question whether 
the different states are temporal manifestations of a single underlying mechanism or 
are they dynamically different, will be vital for a proper modelling of this 
fascinating astrophysical object.

\normalem
\begin{acknowledgements}
KPH and RM acknowledge the financial support from Dept. of Sci. and Tech., Govt. of India, 
through a Research Grant No. SR/S2/HEP - 11/2008.

KPH  acknowledges the hospitality and computing facilities in IUCAA, Pune.
\end{acknowledgements}

\label{lastpage}

\end{document}